\documentclass[twocolumn, prl]{revtex4}
\usepackage{graphicx}
\usepackage{epsfig}
\begin{document}
\title{Intrinsic Heating and Cooling in Adiabatic Processes for Bosons in Optical Lattices}
\author{Tin-Lun Ho and Qi Zhou}
\affiliation{Department of Physics, The Ohio State University, Columbus, OH 43210}
\date{\today}
\begin{abstract}
We show that by raising the lattice  ``adiabatically" as in all current optical lattice experiments on bosons, even though the temperature may decrease initially, it will eventually rise linearly with lattice height, taking the system farther away from quantum degeneracy. This increase is {\rm intrinsic} and is caused by the {\em adiabatic compression} during the raising of red detuned lattices. 
However, the origin of this heating, which is {\em entirely different} from that in the bulk, also shows that one can {\em reverse} the temperature rise to reach quantum degeneracy by {\em adiabatic expansion}, which can be achieved by a variety of methods. 

\end{abstract}

\maketitle

At present, there are worldwide interests to emulate strongly correlated electronic systems using cold atoms in optical lattices. Should these efforts be successful,  we shall have a whole host of  new methods  for studying strongly correlated systems which allow one to vary density, interaction, and dimensionality with great ease. Typically, experiments are   performed in the regime where the quantum gas is in a tight binding band.  
A necessary condition for reaching the strongly correlated regime is that the lattice gas must achieve quantum degeneracy in this narrow band. This can be very challenging, as the temperature for  quantum degeneracy  can be several orders of magnitude  lower than that  in the bulk\cite{osu}.  

In current experiments on bosons, one typically starts with a magnetic trap and then turns on a lattice adiabatically to bring the system into the Mott regime.  How temperature changes in this process is a question of key interest.  While there are theoretical studies\cite{papers,Umass}, there are not yet analytic understanding of the cooling power of this process. 
For a non-interacting gas in a tight binding band, the only energy scale is the hopping integral $t$, which decreases exponentially with lattice height $V_{o}$. By dimensional analysis, the entropy density of a quantum gas in an infinite lattice must be a function of $T/t$\cite{Porto}, where $T$ is the temperature.  At first sight, this seems to furnish a powerful cooling scheme, for $t$ will drop exponentially with increasing $V_{o}$, 
which will force $T$ to do the same to maintain adiabaticity. This, however, does not work in practice.  The entropy density is not only a function of $T/t$ but also $U/t$, where $U$ is the on-site repulsion energy.  As both $t$ and $T$ decrease, $U/t$ becomes so important that it overwhelms the $T/t$ contribution.  Note that interaction effect is not only important in the Mott regime, but also in the superfluid phases close to the quantum critical point where condensate depletion becomes severe.  Thus, even on the superfluid side, 
the ratio $T/t$ can {\em not} be constant  for sufficiently large $V_{o}$. 
   
As a matter of fact, in  an $infinite$ lattice, the temperature of Bose gas {\em must rise } when it is  brought deeper into the Mott regime. This is because the energy gap in the Mott phase is given by $U$, which increases with lattice height $V_{o}$\cite{gap-increases}.  Raising the lattice will therefore make it harder to generate excitations. The only way to keep the entropy constant is then to raise the temperature to counter the rising excitation energy, which unfortunately drives the system  away from quantum degeneracy. We shall call this  ``temperature runaway".

The presence of an harmonic trap, however, has profound effect on the adiabatic processes,  to the point that the bulk effects mentioned above are completely irrelevant. 
In a trap, a lattice Bose gas will have a ``wedding cake" structure consisting of concentric regions of Mott phases. 
Between neighboring Mott regions is a shell of mobile atoms 
(referred to as a ``conducting" shell).  Since boson number fluctuates in these shells, they have much higher entropy density than the Mott regions at temperatures $T<U$, and are the sources of entropy at low temperatures. 
As we shall see, the adiabatic processes in current experiments produces an {\em adiabatic compression} on these conducting shells which again cause a temperature runaway. 
However, by understanding how entropy distributes in the system, it is possible to reverse this ``runaway"   by a simple process, i.e. {\em adiabatic expansion}.  In this way, one can reach down to the degenerate temperature of the order of $t$  even for deep lattices.

{\bf (A). The relevant energy scales:} The potential of an infinite optical lattice is $V_{L}^{}({\bf x}) = V_{o}^{}\sum_{i=1,2,3} {\rm sin}^{2}(\pi x_{i}/d)$, where $d$ is the lattice spacing.  
In deep lattices,  bosons will reside in the lowest band and the system is described by the boson Hubbard model. In grand canonical form, it is ${\cal K} =  \hat{H}_{t} + \hat{H}_{U} - \mu \hat{N}$ $\equiv$ ${\cal K}_{o} + \hat{H}_{t}$,
where $H_{t}=  - t\sum_{\langle {\bf R}, {\bf R'}\rangle} a^{\dagger}_{\bf R}a^{}_{\bf R'}$, $\hat{H}_{U} = \sum_{\bf R} Un_{\bf R}^{}(n^{}_{\bf R} -1)/2$, 
${\cal K}_{o} \equiv  {\cal H}_{U} - \mu \hat{N}$, 
%\begin{equation} H_{t}=  - t\sum_{\langle {\bf R}, {\bf R'}\rangle} a^{\dagger}_{\bf R}a^{}_{\bf R'}, \,\,\,\, \hat{H}_{U} = \sum_{\bf R} Un_{\bf R}^{}(n^{}_{\bf R} -1)/2,  \end{equation}
%\begin{equation}
%{\cal K}_{o} \equiv  {\cal H}_{U} - \mu \hat{N}= \frac{U}{2}\left( \left[n_{\bf R} -\frac{\mu}{U}\right] - \frac{1}{2}\right)^2 - \frac{U}{2}\left( \frac{\mu}{U} + \frac{1}{2}\right)^2,
%\end{equation}
%\begin{equation} H= - t\sum_{\langle {\bf R}, {\bf R'}\rangle} a^{\dagger}_{\bf R}a^{}_{\bf R'} + Un_{\bf R}^{}(n^{}_{\bf R} -1)/2 \label{BH} \end{equation}
%where 
$a^{\dagger}_{\bf R}$ creates a boson at site ${\bf R}$, $n_{\bf R}^{}=a^{\dagger}_{\bf R}a^{}_{\bf R}$, and $\mu$ is the chemical potential. Both the hopping integral $t$ and on-site energy $U$ can be calculated from lattice height $V_{o}$ and recoil energy  $E_{R} \equiv (\pi\hbar)^2 /(2Md^2)$\cite{tU}. 
To give an idea of the energy scales in the Mott regime, we show the values  $E_{G}$, $U$, $t$  and $t^2/U$ in Table 1 for $^{87}$Rb (with scattering length $a_{s}= 5.45nm$) in a lattice with $d=425nm$ for various lattice height, where $E_{G}$ is the lowest band gap in the density of state, and $t^2/U$ is the scale for virtual hopping. 

We see from the Table 1 that for $V_{o}/E_{R}\geq 10$, we have $U >> t>>t^2/U$.  Since the superfluid-insulator transition  is estimated to be  $V_{o}/E_{R}=12\sim13$ for $^{87}$Rb\cite{Bloch}, the condition $U >> t>>t^2/U$ is well satisfied in the region of transition, and is strongly enforced at higher lattice height.  Table 1 shows the great challenge of reaching quantum degeneracy (i.e. the scale $t$\cite{osu}) even for $V_{o}$ as low as $15E_{R}$, and the even greater challenge of reaching temperatures  $\sim t^2/U$. 

\vspace{0.1in}

%\begin{tabular}{||c|c|c|c|c|c|c|c|c||} \hline
%$V_{o}/E_{R}$   & 3   &  5 & 10   &       15 & 20  &  25 & 30 \\  \hline 
%$U$  &   0.0383 & 0.101  &  0.158 &  0.290   & 0.415 &  0.534 & 0.650    & 0.764   \\ \hline
%$t$ &  0.180 & 0.116&  0.0676 & 0.0196 & 0.00670 & 0.0257 & 0.00107 & 0.000477 \\ \hline 
%$t^2/U$ & 0.85 & 0.134  & 0.0290 & 0.00132 & 0.000108 & 0.0000124 &  $1.77 \times 10^{-6}$  & $2.99 \times 10^{-7}$  \\ \hline
%\end{tabular}

\begin{tabular}{||c|c|c|c|c|c|c||}
\hline
$V_{o}/E_{R}$  & 3   &  5 & 10   &       15 & 20 &30 \\
\hline
$E_{G}/E_{R}$ & 0.58   &  1.91 & 4.42   &   6.23 & 7.63 & 9.79 \\
\hline
$E_{G}$ $(nK)$  & 90   &  294 & 678   &   956 & 1171& 1503 \\
 \hline 
 $U$ $( nK)$   & 15.5  &  24.2 &  44.6   & 63.7 &  82.0 & 117.2\\
%$U$ ($0.1 nk$)  &   0.383 & 1.01  &  1.58 &  2.90   & 4.15 &  5.34 & 6.50    & 7.64   \\
\hline
$t$  $(nK)$   & 17.9&  10.4 & 3.01 & 1.03 & 0.39 & 0.073 \\
\hline 
$t^2/U$ $(nK)$   & 20.66  & 4.45 & 0.20 & 0.0166 & 0.00019 &0.00005 \\
\hline
\end{tabular}
%\caption{Table 1. Relevant energy scales of for lattice bosons}

\vspace{0.1in}

{\bf (B). Adiabatic compression:} In current experiments, the optical lattices are constructed from red detuned lasers where atoms are sitting in the region where the laser density is high. The lattice is switched on adiabatically in a magnetic trap
$V_{m}({\bf r}) = M\omega^{2}_{o} r^2/2$, with frequency $\omega_{o}$. However, due to the Gaussian profile of the laser beam, the laser itself  will also produce a confining harmonic potential. When these two  potentials are properly aligned, the frequency $\omega$ of the total harmonic trap $V({\bf r}) = M\omega^{2} r^2/2$  is\cite{gap-increases}
\begin{equation}
\omega^{2} = \omega_{o}^2 + 8V_{o}^{}/(Mw^2),
\label{omega} \end{equation}
where $w$  is the waist of the laser beam.  Thus, as the lattice is switched on, it also provides an {\em adiabatic compression}.  In fact, for strong lattices, the laser contribution can dominate over that of the magnetic trap.  For example, for $V_{o}=12E_{R}$, $E_R=h\times3.2kHz$, $w=130 \mu m$, and $\omega_0=2\pi\times15$Hz, the second term in eq.(\ref{omega}) is already 10 times larger than the first.   As we shall see, this adiabatic compression is the source of temperature runaway in a trap.

{\bf (C). Density profile and entropy distribution:} 
We shall focus on the case $U>>t$, and temperatures $U>>T>0$.
Let us distinguish the cases  $T>t$,  $t>T$, and $T=0$.   ($T$ has the dimension of energy). 

{\bf (C1) $T=0$:}  If $t$ were zero, all sites are decoupled. The eigenstates on each site are number states. 
The ground state then has $\langle n_{\bf R}^{}\rangle=m$, ($m=0,1,2..$), when $\mu$ lies in the interval $(m-1)U<\mu<mU$. In a trap, the density profile can be obtained using local density approximation (LDA) by replacing $\mu$ with  $\mu({\bf r}) = \mu - V({\bf r})$. This leads to a ``wedding cake" structure, (see figure 1),  with sharp steps located at 
\begin{equation}
R_{m}^{} = \sqrt{ 2(\mu -m U)/(M\omega^{2})}, 
\label{Rm} \end{equation}
and  $\mu$ determined by the number constraint, 
\begin{equation}
N=\int {\rm d}{\bf r} n({\bf r})= \frac{4\pi}{3} \sum_{m} (R_{m}/d)^{3} 
%N =  \frac{4\pi}{3} \left( \frac{2}{M\omega^2 \lambda^2} \right)^{3/2} \sum_{m} (\mu - mU)^{3/2}.
\label{Nmu}\end{equation}
The $m$-sum are restricted to $0\leq m <\mu/U$.

Since the energy exciting from $m$ to $(m+1)$ boson state is ${\cal E}_{m+1} - {\cal E}_{m}= mU-\mu$, 
the number states $m$ and $m+1$ are degenerate at $\mu=mU$.   
This degeneracy will be lifted when $t\neq 0$, which introduces fluctuations between neighboring number states. As a result, 
 the sharp steps in fig.1 is rounded off over a shell of width $(\Delta R_{m})^{(o)} \sim (m+1)t/(M\omega^{2}R_{m})$. (See {\bf (C6)} for more discussions). 
  
It is also important to note that {\em in addition to the adiabatic compression, the increase of $V_{o}$ will 
makes the Bose gas more repulsive}, since $U\sim (V/E_{R})^{0.88} \sim V_{o}$\cite{tU}. As a result, $\mu$ also increases  almost linearly as $V_{o}$ (see eq.(\ref{Nmu})) and the radii  $R_{m}$ only has a weak $V_{o}$ dependence. In other words, the trap tightening effect and increase of repulsion, both caused by the rising $V_{o}$, almost cancel each other.  This effect can be verified numerically, and is important for our later discussions.

\begin{figure}
\includegraphics[width=2.8in]{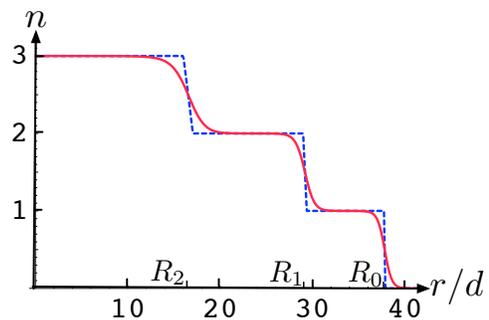}
\caption{Density distribution in a harmonic trap. The density profile of the steps at $T>t$ is described by the Fermi function in eq.(\ref{n}). The system considered has $3.5 \times 10^{5}$ $^{87}$Rb bosons, in a magnetic trap with  $\omega_{o}=2\pi (15Hz)$, $V_{o}/E_{R}=25$, $d=425nm$. }
\end{figure}

{\bf (C2)  $U>>T>>t$:}  In this case, the thermodynamics is again dominated by on-site interaction ${\cal K}_{o}$ with $H_{t}$ as a perturbation. 
It is straightforward to show that  for $\mu \sim mU$, the number density $n(T, \mu)=\langle a^{\dagger}a\rangle/d^3$ is, 
\begin{equation} 
n(T, \mu) = d^{-3}[ m + f ]  + (..),  \,\,\,\,\,  
\label{n} \end{equation}
where $f$ stands for $f(mU-\mu)$,  $f(x) = (e^{x/T} +1)^{-1}$,  and $(..)$ denotes terms of order of ($(t/T)^2$, $e^{-U/T}$) and higher. 

{\em The fact that  the density profile of the step is a Fermi function with width $T$ means that this width can be used as a temperature scale of the system}.  It is also simple to show that the entropy density is
\begin{equation}
s(T,\mu) = d^{-3} \left[ - (1-f){\rm ln}(1-f) - f {\rm ln}f \right] + (..). 
\label{s} \end{equation}
Eq.(\ref{n}) shows that the location of the conducting shell (at $f=1/2$) is still given by  (eq.(\ref{Rm})). 
Applying the Sommerfeld expansion on eq.(\ref{s}) shows that the relation between $\mu$ and $N$ is still given by eq.(\ref{Nmu}) with  corrections  $O(T/U)^2$.  The width of the conducting layer $\Delta R_{m}$ is now given by $\Delta \mu({\bf r}) \simeq T$, or 
\begin{equation}
\Delta R_{m} \simeq T/(M\omega^2 R_{m}). 
\end{equation} 

In figure 2, we have plotted the entropy density as a function of position. One sees that the entropy density $s(r)$ is concentrated in the conducting layers and is essentially zero ($\sim O(e^{-U/T})$) in the Mott region. The maximum value of entropy per site $s({\bf r})d^{3}$ is  ${\rm ln}2 + O((t/T)^2)$. 
It  occurs at $R_{m}$,  reflecting the degeneracy of $m$ and $m+1$ number states at $\mu=mU$. 
The total entropy of the wedding cake structure 
$S_{cake} = 4\pi \int {\rm d} r   r^2  s(T, \mu({\bf r}))$ is then 
$\sim 4\pi \sum_{m} R^{2}_{m}\Delta R_{m} {\rm ln}2 /d^3$, or 
$S_{cake} \sim  \sum_{m} T R_{m}/(m\omega^2 d^{3})$.  
Using Sommerfeld expansion, and with the $(t/T)^2$ correction,  we have\cite{stress} 
\begin{equation}
S_{cake}(T) = \frac{4\pi^3}{3} \frac{T}{M\omega^{2}d^3}\sum_{m} R_{m}\left( 1 + O((t/T)^2, (T/U)^2)\right).
\label{Scake}\end{equation}

\begin{figure}
\includegraphics[width=2.8in]{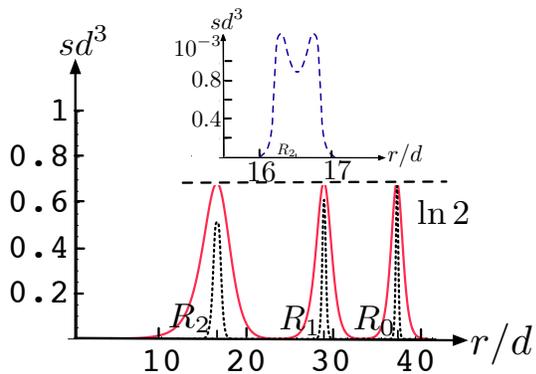}
\caption{Entropy density as a function of position with the same parameters as in fig.1. The solid and dotted curves are for $T=5nK$ and $T=1nK$. The decrease of the peak value of $s$ at smaller $r$ is due to the $(t/T)^2$ and Sommerfeld expansion correction in eq.(\ref{s}). The inset is the entropy density at temperatures $T<t$ discussed in section ${\bf (C6)}$.}
\end{figure}

{\bf (C3) Temperature runaway:}  Eq.(\ref{Scake}) is the origin of temperature runaway. Since $\omega^2$ increases with lattice height (see eq.(\ref{omega})), $T$ must increase to keep the entropy constant.  
The simplest case is a single Mott region, which occurs when $U/2>\mu>0$.  In this case, $R_{0} = \sqrt{2\mu/(M\omega^2)}$, and 
$N=4\pi R_{0}^{3}/(3d^{3})$. The size of Mott regime ($R_{0}$) is independent of $T$ and $\omega$; 
%This  $The number constraint gives $R_{1}= (3N/4\pi)^{1/3}$ which also fixes $\mu$, 
and eq.(\ref{s}) becomes $S_{cake}(T) = (\frac{4\pi^3}{3}) \frac{T}{M\omega^{2}d^2}\left(\frac{3N}{4\pi}\right)^{\frac{1}{3}}$.
In the limit of large $V_{o}$, eq.(\ref{omega}) implies $S_{cake} \propto T/V_{o}$. 
%So $T$ increases linearly as $V_{o}$ in an adiabatic process. 
%This intrinsic heating is detrimental for reaching the correlated regime.  
In general, there are several Mott steps. However, as we have discussed at the end of ${\bf (C1)}$, the sum $\sum_{m}R_{m}$ in eq.(\ref{Scake}) depends weakly on $V_{o}$.  So we again have 
$S_{cake} \propto T/V_{o}$, hence the same temperature runaway.
%The effect of $V_{o}$ on the entropy $S_{cake}$ therefore comes mostly  from $1/\omega^2$, and hence a linear increase of $T$ with $V_{o}$.   

{\bf (C4) Reversing the temperature runaway:} Since the temperature increase is caused by adiabatic compression, the simplest way to reverse it is to reduce $\omega^2$, i.e. an adiabatic expansion.  
Some possible ways are:  (a)  adding a blue detuned laser to generate a repulsive trap (with negative curvature $ -\omega^{2}_{1}$) of variable strength  so that eq.(\ref{omega}) becomes $\omega^2 = \omega_{o}^{2} -  \omega^{2}_{1} + 8V_{o}/mw^2$. (b) Start with a very tight harmonic trap and use a {\em blue}  (instead of red) detuned laser so that  $\omega^2 = \omega_{o}^{2} -  8V_{o}/mw^2$. Expansion can then be achieved by increase $V_{o}$.  (c) Turn on a (magnetic) anti-trap which make the sign of $\omega_{o}^{2}$ negative. This can be achieved by flipping the spin of the boson from low field seeking state to high field seeking. This process is adiabatic as the flipping of spin does not affect the number distribution of the system. (d) Use a laser beam with large width $w$ and reduce $\omega_{o}$.  
The reduction of $\omega^2$ will enable one to reach the lowest possible temperatures within the regime $U>T>t$.  As $T\rightarrow t$, eq.(\ref{Scake}) is no longer valid, and the entropy has to be calculated differently. (See {\bf (C6)}).

{\bf (C5) Condition for cooling:}  Before discussing the case $T<t$, let us illustrate how the initial temperature $T_{i}$ of a Bose gas in a magnetic trap without lattice  is related to the temperature $T_{f}$ of the final wedding cake structure.  The thermodynamics of the a Bose gas in harmonic trap (no lattice) has been studied by  Giogini, Pitaevskii, and Stringari\cite{Sandro} using Popov approximation. For initial temperature $T_{i}<\mu_{o}$, where $\mu_{o}$ is the chemical potential in the center of the trap, their results imply the entropy is 
  \begin{equation} 
S_{i}(T,N) = \frac{7A\zeta(3)}{5\sqrt{2}} \left(\frac{15a_{s} N}{\sigma}\right)^{1/5}\left(\frac{T}{\hbar\omega_{o}}\right)^{5/2}, 
\label{S-harmonic}\end{equation}
where $\sigma=\sqrt{\hbar/(M\omega_{o})}$ and $a_{s}$ is the s-wave scattering length, and $A = 10.6$.

The  temperatures $T_{i}$ and $T_{f}$ are related as $S_{i}(T_{i},N)=S_{cake}(T_{f},N; V_{o})$, where $S_{cake}(T,N; V_{o})$  is obtained by inverting the number relation $N=\int n(T, \mu(r)) = N(T,\mu)$ for  $\mu$  and substituting it into the relation $S_{cake} = \int s(T, \mu({\bf r}))$.   In figure 3, we have represented $S_{i}$ by a black solid line for a system of 
$3.5 \times 10^{5}$ $^{87}$Rb bosons, $\omega_{o}= 2\pi (15Hz)$, $a_{s} = 5.45 nm$. $T_i$ is chosen to be 13nK. The entropies $S_{cake}$ of the final states obtained by raising a red laser lattice to $V_{o}/E_{R} = 15$ and  $30$ in the same magnetic trap  are represented by a dotted and dotted-dash curve respectively. 
At these lattice heights, the system has three Mott layers. 
 The reason that $S_{cake}$ deviates from the linear  dependence on $T$ when $T$ increases is because of the corrections to Summerfeld expansion in eq.(\ref{n}) and (\ref{s}).  
The dashed line with large slope is the entropy of a final state with low (total) trap frequency 
($\omega = 2\pi (15)$Hz),   produced by the expansion schemes mentioned in Section ${\bf (C3)}$. In this case, the system has only one Mott layer.

In the entropy-temperature plot in fig.3,  $T_{i}$ and $T_{f}$ are connected by a horizontal line. Heating (cooling) occurs if $S_{cake}$ lies on the right (left) hand side of $T_{i}$. Temperature runaway  is reflected in the fact that the slope of $S_{cake}(T,N; V_{o})$ decreases with increasing $V_{o}$, and will eventually lie on the right hand side of any $T_{i}$.  For given  $T_{i}$, the  critical $V_{o}$ above  which intrinsic heating occurs is given by $S_{i}(T_{i}, N) = S_{cake}(T_{i}, N; V_{o}^{\ast})$. For our example, this critical $V_{o}^{\ast}$ is about  $25E_{R}$. 
From fig.3, it is clear that if the initial temperature is above $20nK$, raising a red laser lattice will lead to a significant increase in temperature. (See also footnote \cite{Sandro}).  In contrast, if we use a blue laser for the lattice, one can gain an order of magnitude or more reduction in temperature. 

\begin{figure}
\includegraphics[width=2.8in]{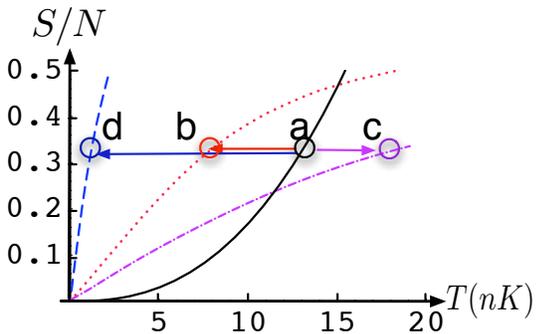}
\caption{Illustration of intrinsic heating and cooling:  Solid curve is $S_{i}$. The dotted and dotted-dashed line are for $V_{o}/R_{R}=15$ and 30 resp.  The system has the same parameters as that in fig.1. Point (a) denotes the initial temperature $T_{i}$. The final temperature for $V_{o}/E_{R}=15$ and 30  are denoted by point (b) and (c). The dashed curve with steepest slope is the total entropy for the adiabatic expansion process with $\omega= 2\pi (15 Hz)$. The final temperature is point (d). For $T_{i}>15 nK$, considerable heating will occur when raising a red laser lattice beyond $V_{o}=15E_{R}$. 
Adiabatic expansion, however, can cool the system considerably. }
\end{figure}

{\bf (C6) $T<t$: } In this case, hopping is not a perturbation. We shall focus on the regime $\mu\sim mU$ where entropy density is highest. This regime is intricate for the following reasons. At $\mu=mU$, the ground state of the Bose Hubbard model  is a superfluid, with transition temperature $T_{c}$ is of the order $(m+1)t$. As the difference $|\mu-mU|$ increases, number fluctuations are suppressed and $T_{c}$ drops. At some point, the falling $T_{c}$ will reach $T$ and drop below it.  To calculate the entropy as a function of $T$ and $\mu$, one must face the complexity of critical phenomenon at some point. While it is possible to calculate the entropy function using quantum Monte Carlo methods, it is desirable to have analytic understanding so as to perform quick estimates, which we do below. 

Near  $\mu\sim mU$, dominant number fluctuations are between the number states $m+1$ and $m$ on each site.  Denoting these two states as a ``pseudo-spin" $|1/2\rangle$ and $|-1/2\rangle$, we have $a_{\bf R} = S^{-}_{\bf R}$, $a^{\dagger}_{\bf R}a^{}_{\bf R} = S^{z}_{\bf R} + m+1/2$; and the Hamiltonian {\cal K} written as\cite{Smitha}, ${\cal K} =   - h\sum_{\bf R} S^{z}_{\bf R} - \sum_{\langle {\bf R, R'}\rangle} [ J_{\perp} {\bf S}^{\perp}_{\bf  R} \cdot {\bf S}^{\perp}_{\bf R'}  
+ J_{z} S^{z}_{\bf  R}  S^{z}_{\bf  R'} ]$
%  \begin{equation}
%{\cal K} =   - h\sum_{\bf R} S^{z}_{\bf R} - \sum_{\langle {\bf R, R'}\rangle} [ J_{\perp} {\bf S}^{\perp}_{\bf  R} \cdot {\bf S}^{\perp}_{\bf R'}   + J_{z} S^{z}_{\bf  R}  S^{z}_{\bf  R'}  ]
%\end{equation}
where ${\bf S}^{\perp}_{\bf  R} = (S^{x}_{\bf  R}, S^{y}_{\bf  R})$, $h$ $ =$ $ \mu - mU + O(t^2/U)$, $J_{\perp}$ $ =$ $ (m+1)t$,  $J_{z}= O(t^2/U)$. 
% and the sum is over nearest neighbors. 
Superfluid order corresponds to $\langle {\bf S}^{\perp}_{\bf R} \rangle\neq 0$.
The Mott phase corresponds to   $\langle S^{z}_{\bf R} \rangle = \pm 1/2$.

At $h=0$, $J_{\perp}\neq 0$, the system at $T=0$ is a ferromagnet in the $xy$-plane (superfluid).  The transition temperature  is $T_{c}\sim J_{\perp}$.    As $|h|$ increases, the ordering spin will tilt away from the $xy$-plane and develop an $S_{z}$ component.  At the same time,   the magnitude of $\langle S^{\perp}_{\bf R} \rangle$ is reduced, and so as $T_{c}$. 
% (i.e. reduced superfluid order due to reduction of number fluctuation caused by asymmetry in excitation energies). 
%For $|h|<J_{\perp}$, the low temperature properties of the system can be calculated u
Using Primakoff-Holstein transformation, it is straightforward to work out the spin wave spectrum and hence 
%The spin wave spectrum is  $\epsilon_{\bf k} = h - 2J_{\perp} \sum_{i} {\rm sin}k_{i}d $. It is then straightforward to evaluate 
the entropy density at temperatures $T<t$ near $\mu\sim mU$, which is  $s_{m}(T,\mu) = C d^{-3} \left( \frac{T}{(m+1)t} \right)^{3} \left( 1 - \frac{|\mu-mU|}{6 t(m+1)}\right)^{-3/2}$, and 
%\begin{equation}
%s_{m}(T,\mu) = C d^{-3} \left( \frac{T}{(m+1)t} \right)^{3} \left( 1 - \frac{|\mu-mU|}{6 t(m+1)}\right)^{-3/2}.
%\label{ss} \end{equation} 
where $C = \sqrt{3} \pi^2 /1620$. 
The fact that $s$ increases as $h$ moves away from 0 reflects the weakening of superfluid order when $h\neq 0$. 

%Since we are at finite temperature, increasing $h$ will cause the superfluid phase to turn normal, 
For $|h|>J_{\perp}$, (or $|\mu - mU|\gg (m+1)t$), the system enters the Mott phase. Since $T< t \sim  J_{\perp}$, the entropy density will reduce to the Mott value (which is essentially zero) over a chemical potential range $\Delta (\mu - V(r)) \sim (m+1)t$, or spatial distance $\Delta R_{m} \sim(m+1)t/(M\omega^{2} R_{m})$.   (See the inset of fig.2). 
The case $|h|\sim J$ is difficult one. Accurate answers will require quantum Monte Carlo treatments. However, with the features of entropy density mentioned above, one can estimate the total entropy to be 
  $S_{cake}\sim 4\pi \sum_{m} R_{m}^2 (\Delta R_{m}) s_{m}$, or $S_{cake}  \sim \frac{4\sqrt{3} \pi^3}{135} \frac{ T^3}{M\omega^2 d^3  t^2} \sum_{m} \frac{R_{m}}{(m+1)^2}.$
This faster drop of $S_{cake}$ with temperature will reduce the cooling power of adiabatic expansion. However, this will take place at temperature scale so low that it is not visible on the scale of fig.3.

%The authors would like express their appreciation for the generosity and hospitality of the Center of Advanced Studes at Tsinghau University, where part of this work is done. 
We would like Randy Hulet and Wolfgang Ketterle for comments on the draft of this paper. 
This work is supported by NSF Grant DMR-0426149 and PHY-0555576.


\begin{thebibliography}{99}
\bibitem{stress} All the figures we show here are exact numerical evaluations of the number constraint $N=\int n(T, \mu(r))$, and total entropy $S_{cake}=\int s(T, \mu(r))$, as well as other quantities. The validity of the leading and sub-leading contributions of the Sommerfeld expansion have all been verified in the temperature range of interest.
\end{thebibliography}
\end{document}